\documentclass[]{raa}
\usepackage{graphicx,times}
\usepackage{natbib}

\providecommand{\bm}[1]{\mbox{\boldmath$#1$\unboldmath}}

\begin{document}

\title{Combining Optical and X-ray Observations of Galaxy Clusters to Constrain Cosmological Parameters}

\volnopage{ {\bf 2010} Vol.\ {\bf 10} No. {\bf XX}, 000--000}
   \setcounter{page}{1}

    \author{Heng Yu
    \inst{}
    \and Zong-Hong Zhu
    \inst{}
    }

\institute{Department of Astronomy, Beijing Normal University, Beijing 100875, China;
{\it zhuzh@bnu.edu.cn }\\
%        {\it yuheng@mail.bnu.edu.cn }\\
\vs \no
        {\small Received [year] [month] [day]; accepted [year] [month] [day] }
}
\abstract{
Galaxy clusters have their unique advantages for cosmology. Here we collect a new sample of 10 lensing galaxy clusters with X-ray observations to constrain cosmological parameters.The redshifts of lensing clusters lie between 0.1 and 0.6, and the redshift range of their arcs is from 0.4 to 4.9. These clusters are selected carefully from strong gravitational lensing systems which have both X-ray satellite observations and optical giant luminous arcs with known redshift. Giant arcs usually appear in the central region of clusters, where mass can be traced with luminosity quite well. Based on gravitational lensing theory and cluster mass distribution model we can derive an Hubble constant independent ratio between two angular diameter distances. One is the distance of lensing source and the other is that between the deflector and the source. Since angular diameter distance relies heavily on cosmological geometry, we can use these ratios to constrain cosmological models. Meanwhile X-ray gas fractions of galaxy clusters can also be a cosmological probe. Because there are a dozen parameters to be fitted, we introduce a new analytic algorithm, Powell's UOBYQA (Unconstrained Optimization By Quadratic Approximation), to accelerate our calculation. Our result proves that this algorithm is an effective fitting method for such continuous multi-parameter constraint. We find an interesting fact that these two approaches are sensitive to $\Omega_{\Lambda}$ and $\Omega_{M}$ separately. Combining them we can get quite good fitting values of basic cosmological parameters: $\Omega_{M}=0.26_{-0.04}^{+0.04}$, and $\Omega_{\Lambda}=0.82_{-0.16}^{+0.14}$ .
}

   \authorrunning{H. Yu  \& Z.-H. Zhu }
   \titlerunning{Combining Optical and X-ray Observations of Galaxy Clusters}
   \maketitle

\section{Introduction}
Cosmic acceleration expansion was discovered initially from
supernovae observations \citep{Riess1998,Perlmutter1999}. Then the cosmic
microwave background (CMB) anisotropy power spectrum
\citep{Spergel2003} confirmed that our universe is nearly flat and
matter density is relatively low. Nowadays such conclusions have
been supported by more precise supernova data \citep{Riess2004,Davis2007,Kowalski2008}
, and CMB observations \citep{Spergel2007,Komatsu2009}. Furthermore, many other independent work, like the light elements abundance from Big Bang
Nucleosynthesis \citep{Burles2001}, the baryon acoustic oscillations(BAO)
detected in SDSS sky survey \citep{Eisenstein2005}, radio galaxies
\citep{Daly2009}, and gamma-ray bursts \citep{Amati2008} etc. also give consistent results. If there is a component with negative pressure, dubbed dark energy, full of our universe, such acceleration can be explained within existing theoretical framework. Then various models were proposed as candidates of dark energy, such as the typical dynamical scalar field quintessence \citep{Caldwell1998}, phantom corrections \citep{Caldwell2002}, a joint quintom scenario \citep{Feng2005} and chaplygin gas \citep{Gorini2005,Zhu2004,2006Zhang} etc. On the other hand there are still many other astronomers doubting the existence of such a strange material. They are trying to find the other way to understand this accelerating universe. Modified Newtonian dynamics(MOND) \citep{Milgrom2001}, Modified Friedmann Equation \citep{Freese2002,Zhu2004b}, Dvali-Gabadadze-Porrati(DGP) mechanism \citep{DGP2000} and so on are all beneficial attempts. But until now none of these models has overwhelming advantages. We need more observational evidence. Besides updating the precision of current data, we also keep trying new probes, for example, galaxy clusters.

Galaxy clusters, as the largest dynamic systems we have known in the universe, are keeping deep imprint of the big bang. Its correlation function provides direct measurement of matter distribution power spectrum. Since the BAO peak has already been found with luminous red galaxies \citep{Eisenstein2005}. When there are enough galaxy clusters, we can do the same measurement in a much larger scale \citep{Borgani2001b}. And their mass distributions at different redshifts can be described by the Press-Schechter function \citep{Press1974}. This relation reflects the linear growth rate of density perturbations. With such connection clusters can provide constraints on matter and dark energy densities \citep{Borgani1999}. What's more, hot gases of galaxy clusters also interact with cosmic microwave background photons and distort its spectrum. Such phenomenon is called the Sunyaev-Zel'dovich effect (SZ effect in short) \citep{Sunyaev1972}. Combining this effect with the observations of corresponding clusters' X-ray luminosity, we can measure the Hubble constant for a certain cosmology or give a rough estimate of cosmological parameters \citep{Mason2001,Reese2002,Schmidt2004,Jones2005,Bonamente2006,2004Zhu2}. For more detailed reviews we suggest \citet{Rosati2002}, \citet{Voit2005} and \citet{Borgani2006}.

The two methods adopted here are based on the physical structure and properties of individual clusters. They can also give a good estimate of cosmological parameters as you will see. The first one comes from strong lensing arcs. From X-ray luminosity and temperature we can model mass distribution of a cluster's mass halo. Giant arcs generated by high concentrated gravity of a galaxy cluster are perfect indicators of its surface mass density. Then we can derive an observational value to constrain cosmological models. This method was first used by \citet{Sereno2002}, and improved in \citet{Sereno2004}. We collect a new data set from literature and an online database BAX. Several effective criteria are introduced to rule out improper clusters. In the end 10 clusters are selected to make a new sample. This new set can give an interesting result compared with the last one supplied by \citet{Sereno2004}. We will discuss it in detail in section 2; The second way is based on the assumption of constant x-ray gas mass fraction. This method is developed by \citet{Allen2001,Allen2004,Allen2008} and has been proved to be effective. Due to the complex physical mechanism involved, there are many parameters to fit. To search multi-parameters space more effectively, we introduce a new optimization algorithm UOBYQA (Unconstrained Optimization By Quadratic Approximation) \citep{Powell2002} to marginalize external parameters. This algorithm has been widely accepted in mathematical field. This should be its first application in cosmology. The algorithm is summarized in section 3; The combined analysis and discussions are presented in the last section.

\section{Lensing Cluster}
\subsection{Theoretical foundation}
Gravitational lensing is one of successful predictions of general relativity. After it is confirmed by QSO observation in 1979 \citep{1979Natur}, \citet{Paczynski1981} tried to use these lensing images as indicators to estimate cluster mass and constrain cosmological constant. But it is really hard to find the deflectors in lensing events of point-like sources. So this approach was seldom used until recent years \citep{Futamase2001}. Later on, giant arcs around galaxy clusters were discovered in clusters A370 and Cl2224 \citep{Lynds1986}. They can also be used to constrain clusters' projected mass, and clusters are easier to observe. \citet{Breimer1992} did their pioneering work. They estimated the virial mass of cluster A370 with nearly 30 member galaxies' velocities, compared with lensing condition and got an original estimate of cosmological parameters. But member galaxies are discrete and velocity dispersion of distant objects is usually hard to obtain. Then \citet{Sereno2004} use continuous X-ray luminosity instead. When a galaxy cluster is relaxed enough, the pressure of its hot gas can balance its self-gravity. In this case, we can use hydrostatic isothermal spherical symmetric $\beta$-model \citep{Cavaliere1976} to describe the intracluster medium(ICM) density profile:

\begin{eqnarray}\label{nr}
n_{e}(r)=n_{e0}\left( 1+r^{2}/r^{2}_{c}\right) ^{-3\beta_{X} /2},
\end{eqnarray}
where $n_{e0}$ is the central electron density, $\beta_{X}$ describes the slope, and $r_{c}$ stands for the core radius. Assuming all the gases have the same emissivity of hot bremsstrahlung radiation, in other words, isothermal (with the temperature $T_{X}$), the gravity of relaxing cluster and its gas pressure should balance each other according to the hydrostatic equilibrium condition. With the approximation of spherical symmetry we can estimate mass distribution with gas density, which comes from luminosity fitting result. The cluster mass profile can be given in this form:

\begin{eqnarray}\label{mass}
M(r)=\frac{3k_{B}T_{X}\beta_{X} }{G\mu m_{p}}\frac{r^{3}}{r^{2}_{c}+r^{2}},
\end{eqnarray}
$k_{B}$ is the Boltzmann constant, $m_{p}$ is the proton mass, and $\mu$ is the mean molecular weight, which is 0.6 usually \citep{Rosati2002}. Then we can get the projected surface mass density. Combining the result with the critical surface mass density of lensing arcs \citep{Schneider1992}, a Hubbel constant independent ratio can be expressed by several observational parameters:
\begin{eqnarray}
\frac{D_{ds}}{D_{s}}\Big|_{obs}=\frac{\,\mu m_{p}c^{2}}{6
\pi}\frac{1}{k_{B}T_{X}\beta_{X}}\sqrt{
\theta_{t}\!^{2}+\theta_{c}\!^{2} },
\end{eqnarray}
where $T_{X}$, $\beta_{X}$, and $\theta_{c}$ are all come from
the X-ray data fitting. $\theta$ is a dimensionless angular variable, radius divided by angular diameter distance to the cluster ($r/D_{d}$). Under flat Friedman-Walker metric,the angular diameter distance between an observer at $z_{d}$ and a source at $z_{s}$ is not equal to $D_{s}-D_{d}$. It should be integrated from $z_{d}$ to $z_{s}$ as below :
\begin{eqnarray}
\label{inted}
D_{ds}=\frac{c}{H_{0} (1+z_{s})}\int_{z_{d}}^{z_{s}} \frac{dz}{E(z)}.
\end{eqnarray}
The position of tangential critical curve $\theta_{t}$ is usually deemed equal to observed arc position $\theta_{arc}$. Considering the deviation of extended lensing source position, deflecting angle has a slight difference with arc radius angle, $\theta_{t}=\epsilon\theta_{arc}$. The correction factor is $\epsilon=(1/\sqrt{1.2})\pm0.04$ \citep{Ono1999}. Then the $\chi^{2}$ test can be carried out between observational data and theoretical models.

\subsection{Data selection}

The number of the clusters with arcs is still very limited, and just a small part of these arcs have known redshift. At first, we refer to the sample of \citet{Sand2005} to look for arcs with redshift. Their catalogue contains 104 tangential arcs from 128 clusters, and only 58 arcs from 27 clusters have redshift values. When there are several arcs from the same source (with the same redshift), we prefer to choose the farthest one in general. Because strong lensing arcs usually happen in the very central part of galaxy clusters, giant arcs are always close to the core of clusters. While the beta model brightness profile usually has better fitting results at outskirts. So we choose the farther part to avoid possible substructures and decrease the fitting deviation of $\beta$ model. Then arc H5b (26.3 arcsec) of A2390 is adopted instead of H5a (20.7 arcsec). For the arcs from different sources around one cluster, we treat them dependently. They will all be adopted as long as they can satisfy our criteria given below.

Second stage, redshifts and temperatures of these galaxy clusters can be searched out directly from online databases, such as CDS (The Strasbourg astronomical Data Center) or NED (NASA/IPAC Extragalactic Database) with their full name. Here we choose a new database established especially for X-ray galaxy clusters -- BAX\footnote{BAX: http://bax.ast.obs-mip.fr/ , same with the other database like Simbda, NED, etc., only accept clusters' full name like RXJ1347.5-1145. The abbreviations listed here is only for convenience, can not be used to search directly.}. It provides detailed information of clusters. Following referred literature given by BAX we can get fitting parameters $\beta$ and $\theta_{c}$. To minimize systematic error brought by different work and systems, it is better to limit data sources within a few articles. In this paper, we use the fitting result of Chandra \citep{Bonamente2006}. For the clusters not presented by that paper, we refer to the catalogue\footnote{Ota's catalogue wasn't contained in the article, but was put on her own website.} of \citet{Ota2004}, which is based on the observations of ROSAT and ASCA satellites. The clusters inherited from \citet{Sereno2004} are all updated in this way.

\begin{table*}[ht]
\caption{Sample of Lensing Galaxy Clusters with X-ray Observations}
\label{list}
\begin{center}
\begin{tabular}{lcccccccccc}

Cluster & Arc & z$_d$ & z$_{arc}$ & $\theta_{arc}$(") & $kT_{X}(keV)$ &
$\beta_{X}$ & $\theta_c$(") & n($\rho_{0}(z)$) & ref \\
\hline
3C220.1 & A1 & 0.61 & 1.49 & 20.3 & 5.56$\pm$1.38 & 0.84$\pm$0.45 & 8.1$\pm$4.2 & 8537 & 1 \\
Abell 2390 & H5b &  0.228   & 4.05 & 26.3 & 9.35$\pm$0.15 & 0.46$\pm$0.01 & 12.1$\pm$1.4 & 59463 & 1 \\
Abell 2667 & A1 & 0.226 & 1.034 & 14.7 & 6.15$\pm$0.61 & 0.52$\pm$0.01 & 13.4$\pm$0.8 & 65462 & 1 \\
MS 0451.6-0305  & A1 & 0.550 & 2.91 & 31.8 & 10.4$\pm$0.7 & 0.767$\pm$0.018 & 33.5$\pm$1.2 & 4177 & 2 \\
MS 1512.4 & cB58 & 0.372 & 2.72 & 5.1 & 3.39$\pm$0.4 & 0.54$\pm$0.06 & 8.3$\pm$2.4 & 52805 & 1 \\
MS 2137.3-2353 & A01 & 0.313 & 1.501 & 15.5 & 4.96$\pm$0.11 & 0.6$\pm$0.04 & 8.3$\pm$1.5 & 38603 & 1 \\
PKS 0745-191 & A & 0.103 & 0.433 & 19.2 & 7.97$\pm$0.28 & 0.52$\pm$0.01 & 16.4$\pm$0.9 & 271090 & 1 \\
Abell 68 & C0c & 0.255 & 1.6 & 8.0 & 10.0$\pm$1.1 & 0.721$\pm$0.035 & 49.6$\pm$3.6 & 17867 & 2 \\
CL0024.0 & E & 0.391 & 1.675 & 4.0 & 4.38$\pm$0.27 & 0.41$\pm$0.03 & 11.1$\pm$4.1 & 31586 & 1 \\
MS 2053.7  & AB &  0.583 & 3.146 & 15.1 & 4.7$\pm$0.5 & 0.639$\pm$0.033 & 15.3$\pm$1.6 & 6324 & 2 \\
\hline

\end{tabular}\\
Reference: 1.\citet{Ota2004}; 2.\citet{Bonamente2006}
\end{center}
\end{table*}

Then we collect about 20 clusters with all necessary parameters. But not all of them are approximately isothermal, spherical symmetric, or even hydrostatic equilibrium. We must check them carefully with criteria to eliminate the arcs generated by unrelaxed clusters \citep{Smith2003}.

\subsection{Criteria}
A static cluster should have regular morphology both in optical and
X-ray band. In optical band they usually present spatial symmetry
and regular shape. In X-ray band sharp central surface brightness and
regular elliptical isophotes are both common characters of dynamically
relaxed clusters. But such description rely on the resolution of equipment greatly. We need more concrete standards.

\begin{figure}
\includegraphics[width=1\textwidth]{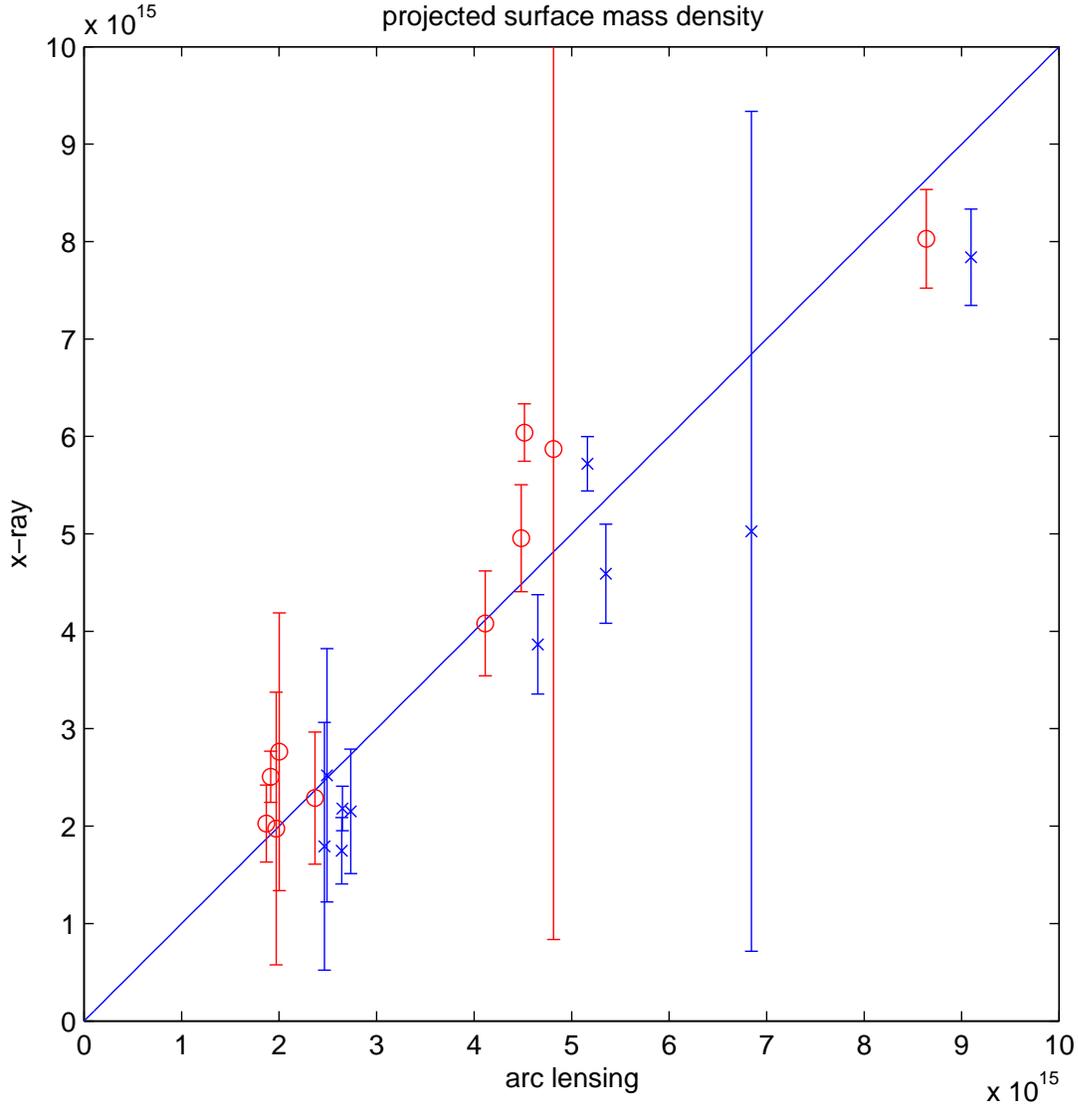}
\caption{\label{rho}Relationship between lensing surface mass density($M_{\odot}/Mpc^{2}$) and projection surface density from X-ray model.Blue asterisks come from standard model ($\Omega_{M}=0.3$,$\Omega_{\Lambda}=0.7$);and red circles are calculated under $\Omega_{M}=0$,$\Omega_{\Lambda}=1$.}

\end{figure}

Weak gravitational lensing has been used to test X-ray mass distribution many years ago. \citet{Mahdavi2008} confirmed that they were consistent with each other at inner part. Although he used NFW dark matter halo model \citep{NFW}, the results make sure that hydrostatic equilibrium is applicable within radius $r_{2500}$(which means the mean mass density is 2500 times of the universe critical density at that redshift). We can find from Table \ref{list} that all our arcs appear deep inside this region. To avoid possible big disagreement of inner mass estimation, we also compare lensing mass with theoretical predictions to check our mass model. According to the strong lensing equation of isothermal sphere, the projecting mass is:
\begin{eqnarray}
\Sigma_{ob}= \frac{c^{2}}{4\pi G} \frac{D_{s}}{D_{d} D_{ds}}\sqrt{
\frac{\theta_{t}^{2}}{\theta_{c}^{2}}+1}.
\end{eqnarray}
And from equation (\ref{mass}), we can derive a theoretical typical surface density, which is:
\begin{eqnarray}
\Sigma_{th}= \frac{3}{2G \mu m_{p}}
\frac{k_{B}T_{X}\beta_{X}}{\theta_{c}}\frac{1}{D_{d}}.
\end{eqnarray}
Mass values from above two equations should be consistent with each other for a spherical relaxed cluster as shown in Fig.\ref{rho}. Because this test is not model independent, we don't take it as a selecting tool, just use it to examine the data with the best fit model.

\begin{figure}
\includegraphics[width=1\textwidth]{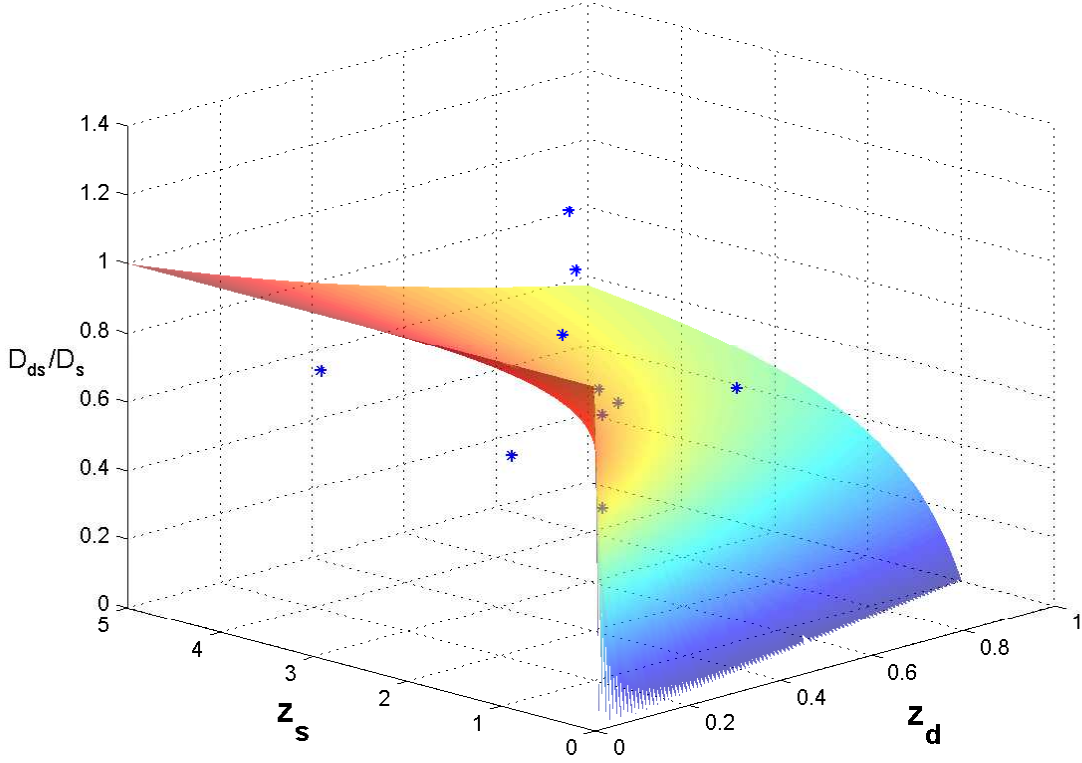}
\caption{\label{3DHubble}3-D Hubble diagram distribution of sample galaxy clusters.X axis is the
redshift of deflectors, Y the arc, and Z is the second criteria $D_{ds}/D_{s}$, the reference surface comes from our best-fit model below. The intersecting line between fitting surface and the bottom of the box is $z_{d}=z_{s}$. And the jagged edge is merely caused by programme.}
\end{figure}

In equation (\ref{inted}), distance between the len and the source (integrated from $z_{d}$ to $z_{s}$) should be always smaller than that between the arc source and the observer (integrated from 0 to $z_{s}$), although the angular distance is not monotonically ascending with redshift. So these clusters should satisfy:  $D_{ds}/D_{s}|_{obs}<1$. In Fig.\ref{3DHubble} we can see this more clearly. The points above the box indicate corresponding clusters are not self-consistent. They need more precise descriptions for cosmological test. Unfortunately this condition rules out half of selected lensing arcs.

And we also discard the arcs whose positions are too far from characteristic radius ($\theta_{arc} > 3 \theta_{c}$). Such cluster has a relatively small core radius and a much bigger arc radius. In this case, even X-ray observations can not trace the matter at the region where lensing arcs existing. The extrapolation results of the model for inside region is not reliable any more. They may influence constraint result greatly. For example, the most luminosity cluster in the sky RXJ1347 is still experiencing merging process \citep{Allen2004,Ota2008}. It will cause really serious deviation when counted in. In fact it is in charge of the high $\Omega_{\Lambda}$ value (about 1) in the paper of \citet{Sereno2004}. This may also explain the systematic bias in the articles of \citet{Breimer1992} and \citet{Sereno2002}. A similar situation also happens to C arc (38.1 arcsec) of A2390. So only an inner arc H5b is adopted.

At last we get a sample of 10 clusters listed in Table \ref{list}. The redshifts of these clusters rang from 0.1 to 0.6, and the farthest arc is H5b of A2390 z=4.05. All these arcs appear in very central region. The average surface density within the arcs are listed at n column with the unit of the critical universe density at corresponding redshift. And data for cosmological fitting are shown in Table \ref{cosmo}. Errors are calculated with error propagation equation.

\begin{table}
\caption{Values of $D_{ds}/D_{s}$ from Eq.(3)}
  \label{cosmo}
  \begin{center}
  \begin{tabular}{ccc}
  Cluster & $D_{ds}/D_{s}$ & $\sigma(D_{ds}/D_{s})$ \\
  \hline
  3C220.1& 0.611 & 0.530 \\
  A2390 & 0.737 & 0.053 \\
  A2667 & 0.837 & 0.124 \\
  MS0451 & 0.785 & 0.087 \\
  MS1512 & 0.734 & 0.330 \\
  MS2137 & 0.778 & 0.105 \\
  PKS0745 & 0.818 & 0.065 \\
  A68 & 0.982 & 0.225 \\
  CL0024 & 0.919 & 0.430 \\
  MS2053 & 0.968 & 0.209 \\
  \end{tabular}
  \end{center}
\end{table}

\section{X-ray gas mass fraction}
There is another way to connect physical characters of galaxy clusters to cosmological parameters. As mentioned above, in hydrostatic equilibrium the X-ray gas mass distribution and the total mass profile which balance it can both be derived from the surface brightness profile \citep{White1991}. Because of the large scale of clusters, their matter content can be taken as a fair sample of the whole Universe \citep{1993White}. Since the baryonic fraction of clusters are sensitive to angular diameter distance. \citet{1996Sasaki,1997Pen} first applied this to cosmological test. \citet{Allen2001} assume that the X-ray gas mass fraction within $r_{2500}$ is a constant. Then the angular distance of clusters can be derived based on reference cosmological model \citep{Allen2004,Zhu2004a,Allen2008}. In reference cosmology $h=H_{0}/100kms^{-1}Mpc^{-1}$,$\Omega_{\Lambda}=0.7$ and $\Omega_{m}=0.3$, we have
\begin{eqnarray}
f_{gas}^{\Lambda CDM}(z)=\frac{K A \gamma b(z)}{1+s_{z}}\frac{\Omega_{b}}{\Omega_{m}}\left[\frac{d_{A}^{\Lambda CDM}(z)}{d_{A}^{mod}(z)}\right]^{1.5},
\end{eqnarray}
where K is a calibration constant, A equals to $(\theta_{2500}^{\Lambda CDM}/\theta_{2500})^{\eta}$, $\gamma$ stands for non-thermal pressure in the clusters; b(z) is the depletion factor with the expression of $b_{0}(1+\alpha_{b}z)$; s(z) models the baryonic mass fraction in stars, also expressed as $s_{0}(1+\alpha_{s}z)$. Two weak priors are also needed here: Hubble parameter $h=0.72 \pm 0.24 $ and mean baryon density $\Omega_{b}h^{2} = 0.0214 \pm 0.006 $ \citep{Allen2008}.

Obviously, so many parameters aren't easy to calculate with normal methods. \citet{Allen2008} used the popular MCMC (Markov Chain Monte Carlo) programme CosmoMC \citep{Lewis2002}. For such a nonlinear or non-derivable function, stochastic process can generate right distribution, but it need a large point set to draw smooth contour lines. And we usually don't need to obtain all parameters' bias at the same time. In this case we only care about cosmological parameters. So we can just constrain 2 parameters each time and marginalize the rest. In actual calculation we assemble the parameters K, $\gamma$, $b_{0}$, $S_{0}+1$ into one factor. Marginalizing them together will not affect our final results. Then we use grids to generate reference points on parameters' phase space. It is easy to get contour lines with them. We just need to search for the most optimistic value in other parameters' space to accomplish marginalization. Considering all the physical processes are still continuous, their functions have definite slope, we adopt a new analytical algorithm--direct search algorithm. Its main motivation is using numerical finite difference to approach directional derivatives. Here we use a ready-made program module CONDOR (COnstrained, Non-linear, Direct, parallel Optimization using trust Region method for high-computing load function)\footnote{It is available at $http://www.applied-mathematics.net$} \citep{Frank2005}. It is based on the Powell's UOBYQA algorithm \citep{Powell2002}.

With a sample of 42 galaxy clusters \citep{Allen2008}, we obtain the marginalized 68 percent confidence limits of $\Omega_{M}=0.26\pm0.04$ and $\Omega_{\Lambda}=0.9_{-0.18}^{+0.14}$, minimum $\chi^{2}=41.8$. Contours are shown with dash lines in Fig.\ref{comb3}. This result is consistent with the results of \citet{Allen2008} very well, which is $\chi^{2}=41.5$, $\Omega_{M}=0.27\pm0.06$ and $\Omega_{\Lambda}=0.86\pm0.19$.  Contour lines are showing with dashed lines in Fig.\ref{comb3}.

\section{Result and Discussion}
\begin{figure}
\includegraphics[width=1\textwidth]{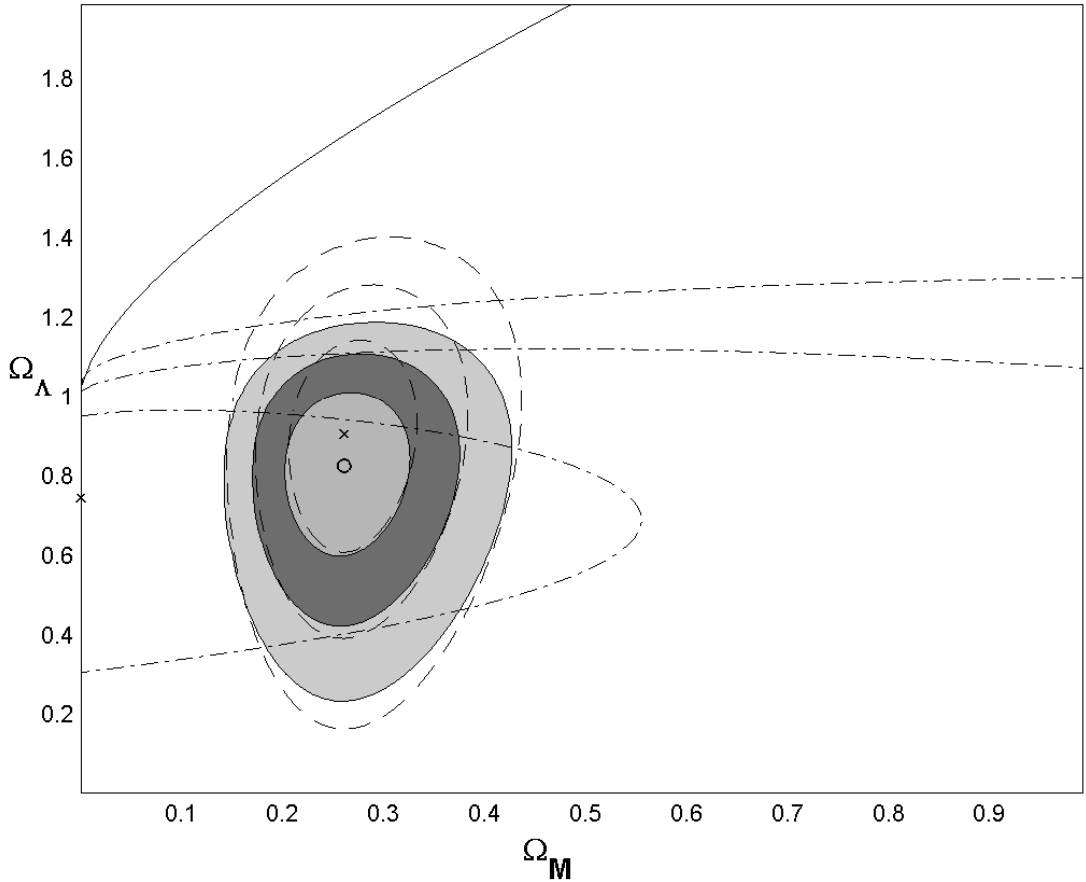}
\caption{Dash line gives the result of X-ray gas mass fraction. While dash dot line comes from lensing cluster. Two "x" simbols  give their best fitting values seperately. The shaded regions show the 1,2,and 3 $\sigma$ confidence regions of combining constrain, corresponding to $\Delta\chi^{2}$ values of 2.30,6.17,and 11.8. The small circle inside is the best fitting point: $\Omega_{M}=0.26_{-0.04}^{+0.04}$ and $\Omega_{\Lambda}=0.82_{-0.16}^{+0.14}$. }
\label{comb3}
\end{figure}

Our sample is not big enough to make a precise constraint to the equation of state. So we use a simple cosmological model $E(z)=\sqrt{\Omega_{M}(1+z)^3 + \Omega_{\Lambda} + (1-\Omega_{M}-\Omega_{\Lambda})(1+z)^{2}}$ . Because these two methods use different cluster sample, we fit them separately and sum their $\chi^{2}$ up to get the final fitting results. As can be seen in Fig.\ref{comb3}, dash lines give the result of X-ray gas mass fraction, and dash dot contours come from lensing clusters. The shaded region shows the combining constrain. The small circle inside gives the best fitting values: $\Omega_{M}=0.26_{-0.04}^{+0.04}$ and $\Omega_{\Lambda}=0.82_{-0.16}^{+0.14}$ at 68\% C.L. . These results are in agreement with the basic facts we know from other observations.

When combining results of these two methods, we find that these two constraining methods coming from clusters are sensitive to different cosmological parameters separately. Their contour regions are nearly orthogonal. The angular diameter distance ratio estimated from strong lensing arcs relies on $\Omega_{\Lambda}$ sensitively. The best fitting value is $\Omega_{\Lambda}=0.74_{-0.36}^{+0.18}$. It's better than the result from constraint of gas fraction. But it can not give an significant constraint to $\Omega_{M}$. On the other hand, the gas fraction of clusters is more effective in constraining $\Omega_{M}$. So when we combine these two methods together, unified constraint is better than both of them. Such results are still preliminary. Due to the complexity of clusters themselves, there is still a lot of work to do to improve these methods.

Common radial temperature distribution of clusters make it difficult to use isothermal approximation. But in the small scale within arcs temperatures won't change steeply even for cool-core cluster. The clusters which have substructures or even are experiencing merging can not be discribed simply. At least we can rule them out by our criteria.  The X-ray observations of our new lensing cluster sample come from three different satellites: ROSAT, ASCA \citep{Ota2004} and CHANDRA \citep{Bonamente2006}. The differences of equipments may also cause systematical errors which are hard to estimate. Considering previous two telescopes are comparatively out of time, fitting results from CHANDRA or XMM may give better results.

These two methods use two different cluster sample sets with some common clusters (A2390, MS2137 etc.). The deviation of different observations may also be counted in indirectly. So it's necessary to select a unified set to share the same clusters and observational parameters. On the other side, the mass profile model used by these methods are different. The strong lensing approach uses the $\beta$-model to fit luminosity profile of a cluster, while the gas fraction method uses the NFW model to describe dark matter halo. As we have seen in the Table \ref{rho}, lensing arcs usually appear in the central region of clusters, where the NFW model can usually give a better fitting result than that of the isothermal sphere model \citep{Comerford2006,Schmidt2007}. If we can unify the two methods with the same mass profile model, results will be more convincing.

Compared with the other cosmological observations, our clusters sample
is really small, and its range of redshift is also limited. According to \citet{Yamamoto2001}, a data set containing more than 20 clusters can constrain dark energy equation of state more precisely. It won't take a long time to achieve that goal. There are many giant arc survey projects proceeding \citep{Gladders2003,Hennawi2008}. The number of new discovered arcs is increasing rapidly. Their redshift measurement is only a matter of time. And next generation X-ray telescope, eg. International X-ray Observatory (IXO) \citep{2010IXO}, extended ROentgen Survey with an Imaging Telescope Array (eRosita) \citep{2010eRosita} and the Wide Field X-ray Telescope (WFXT) \citep{2010WFXT}, will carry out new surveys more precisely in a much larger field. Future observations will definitely enlarge our set and make these methods more powerful.

\normalem
\begin{acknowledgements}
This work was supported by the National Science Foundation of China under the Distinguished Young Scholar Grant 10825313 and by the Ministry of Science and Technology national basic science Program (Project 973) under grant No. 2007CB815401. We are grateful to Zi Xu for her constructive suggestions on
optimization algorithms and introducing the powerful program CONDOR.
We also thank Paolo Tozzi, Xiang-Ping Wu and Xing Wu for their helpful comments.
\end{acknowledgements}

\end{document}